# Near-Infrared plasmon induced hot electron extraction evidence in an indium tin oxide nanoparticle / monolayer molybdenum disulphide heterostructure

Michele Guizzardi[1], Michele Ghini[2], Andrea Villa[1], Luca Rebecchi[2,3], Qiuyang Li[4,5], Giorgio Mancini[6], Fabio Marangi[1], Aaron M. Ross[1], Xiaoyang Zhu[4], Ilka Kriegel[2], Francesco Scotognella[1]

1 Dipartimento di Fisica, Politecnico di Milano, piazza Leonardo da Vinci 32, 20133 Milano, Italy
2 Functional Nanosystems, Istituto Italiano di Tecnologia, via Morego 30, 16163, Genova, Italy
3 Dipartimento di Chimica e Chimica Industriale, Università degli Studi di Genova, Via Dodecaneso 31, 16146 Genova, Italy
4 Department of Chemistry, Columbia University, 3000 Broadway , Havemeyer Hall · New York, NY 10027, USA
5 Department of Physics, University of Michigan, 450 Church Street, Ann Arbor, MI 48109-1040, USA
6 Smart Materials, Fondazione Istituto Italiano Di Tecnologia, Via Morego 30, 16163, Genova, Italy

**Abstract**

In this work, we observe plasmon induced hot electron extraction in a heterojunction between indium tin oxide nanocrystals and monolayer molybdenum disulphide. We study the sample with ultrafast differential transmission exciting the sample at 1750 nm where the intense localized plasmon surface resonance of the indium tin oxide nanocrystals is and where the monolayer molybdenum disulphide does not absorb light. With the excitation at 1750 nm we observe the excitonic features of molybdenum disulphide in the visible range, close to the exciton of molybdenum disulphide. Such phenomenon can be ascribed to a charge transfer between indium tin oxide nanocrystals and monolayer molybdenum disulphide upon plasmon excitation. These results are a first step towards the implementation of near infrared plasmonic materials for photoconversion.

Alternative ways to produce energy from the Sun also include plasmon induced hot electron extraction based solar cells [1,2]. In 2016, Reineck et al. have demonstrated photon-to-electron conversion efficiency by employing a solar cell with an interface between gold nanoparticles and titanium dioxide. Visible light excites the gold nanoparticles generating hot electrons. Such hot electrons are higher in energy with respect to the bottom of the conduction band of titanium dioxide. Thus, hot electron transfer has been observed in this heterojunction [3]. Plasmon-induced hot electron extraction has been observed also in Au/p-GaN heterostructures [4].

With materials that show the plasmonic resonance in the infrared it is possible to observe plasmon-induced hot electron extraction in the infrared [5,6]. In recent years the attention on plasmons in heavily doped semiconductor nanocrystals has increased. Doping levels around $10^{20-21}$ cm$^{-3}$ place their plasmon resonances in the near infrared. Of particular interest are transparent conducting oxide nanocrystals. Doping control, doping placement and the variety of different dopants and host lattices to choose gives a versatile tool box to produce materials that cover desired spectral ranges [7–9]. Also in indium tin oxide, synthesis techniques have allowed to improve the plasmon resonance quality through indium oxide shell growth [10–12,9]. Recently, hot electron extraction has been demonstrated in a heterostructure between indium tin oxide nanocrystals and tin oxide nanocrystals [13] and between fluorine indium co-doped cadmium oxide nanocrystals and rhodamine 6G dyes [14].

Herein, we show plasmon-induced hot electron extraction in an indium tin oxide nanocrystal / monolayer molybdenum disulphide structure. By exciting the heterostructure at 1750 nm, where molybdenum disulphide is not absorbing light and indium tin oxide shows a strong absorption, we observe the excitonic feature of molybdenum disulphide. We ascribe this phenomenon to plasmon-induced hot electron transfer between indium tin oxide nanocrystals and monolayer molybdenum disulphide.

*Sample preparation*: MoS$_2$ monolayer sample was prepared using the reported method with slight modifications [15]. We first deposited a 150 nm gold film on a Si wafer with e-beam evaporation (0.05 nm/s), then spin-coated polyvinylpyrrolidone (PVP) solution (Sigma Aldrich, mw 40000, 10% wt in ethanol/acetonitrile wt 1/1) on the gold film (1500 rpm for 2 min, acceleration with 500 rpm/s) and heated it at 150 °C for 2 min. Next, we put the heat release tape onto PVP/gold surface to peel off the gold from the Si wafer and pressed the gold surface onto a MoS$_2$ single crystal and peel off a monolayer MoS$_2$. The MoS$_2$ monolayer on gold was then transferred onto a fused silica substrate. We first removed the heat release tape by heating the tape/PVP/gold/MoS$_2$ on fused silica substrate at 130 °C for 5 min, then removed the PVP layer by water-soaking for 3 hours, and finally removed the gold film by gold etchant (2.5g I$_2$ and 10g KI in 100 mL deionized water). The MoS$_2$ monolayer on fused silica was washed by water and isopropanol, then dried by a nitrogen gun.

Indium tin oxide (ITO) nanocrystals (NCs) were synthesized in the following procedure. Indium(III) acetate (CAS:25114-58-3), Tin (IV) acetate (CAS:2800-96-6), oleic acid (technical grade, 90% purity, CAS:112-80-1), oleyl alcohol (technical grade, 85% purity, CAS:143-28-2) were purchased from Sigma Aldrich. As first step, a 500 mL three neck round flask was filled with 208 mL of oleyl alcohol and left at 150 °C to degas for 3 hours under a flux of nitrogen. Indium and tin precursors were added, along with 32 mL of oleic acid, to a 100 mL three neck, round bottom flask. Under stirring, the flask content was degassed for three hours under a nitrogen flux, allowing tin and indium oleates to form. After degassing, the flask with oleyl alcohol, which will act as the reaction vessel, was kept under a flux of 0.130 L/min of nitrogen and heated to 290°C. Indium and tin precursors were transferred in a syringe and injected in the hot oleyl alcohol using a syringe pump with an injection rate of 4.8 mL/min. After 15 minutes since the injection ended, resulting in NCs with an average diameter of 13 nm and 10.8% doping level (Sn/tot). The solution was then centrifuged at 5540 G for

10 minutes, using ethanol as antisolvent. The supernatant was discarded, the material redispersed in hexane, ethanol added and centrifuged again using the same parameters. Finally, the NCs were stored in octane.

The $MoS_2$-ITO 2D-0D hybrid was prepared by spin-coating 30 μL of the 10 mg/mL nanocrystal solution at 2000 rounds per minute (rpm) for 45 seconds, with a ramp time of 10 seconds, over the $MoS_2$ sample. The hybrid was then heated at 300 °C for 1 h in inert atmosphere to improve the film conductivity.

*Sample characterization:* Transmission electron microscopy (TEM) was performed to structural characterized ITO NCs and determine the size distribution. The images were acquired by depositing the NCs on lacey carbon grids supported by a copper mesh and using a JEOL JEM-1400Plus operating at 120 kV. Statistical analyses on the acquired images were carried out by using ImageJ software (NIH) and OriginPro software.

Inductively coupled plasma mass spectrometry (ICP-OES) was performed to estimate the doping level of the ITO NCs. The elemental analysis was carried out on an iCAP 6000 Series ICP–OES spectrometer (Thermo Scientific). The NCs were dissolved in aqua regia [HCl/HNO3 3:1 (v/v)] and left overnight at room temperature. Then, Milli-Q grade water (18.3 MΩ cm) was added to the sample. The resulting solution was filtered using a polytetrafluorethylene membrane filter with 0.45 μm pore size.

Photoluminescence and Raman spectroscopy measurements were performed by using a Renishaw microRaman inVia 1000 with a 50× objective (N.A. = 0.75) and an excitation wavelength of 514.5 nm. Raman spectra were collected from 41 $cm^{-1}$ to 1412 $cm^{-1}$ with a resolution of 1.5 $cm^{-1}$, while PL spectra were collected from 592 nm to 767 nm with a resolution of 0.17 nm. Hyperspectral images were generated by scanning the sample and collecting Raman and PL spectra for each spatial position and then analysed with Python-based dedicated code.

Fourier Transform Infrared (FTIR) Spectroscopy was conducted on a Vertex 70v vacuum spectrometer (Bruker) in transmission configuration on both the $MoS_2$ and the $MoS_2$-ITO hybrid samples. Spectra were recorded in the wavenumber region of 6000 to 600 $cm^{-1}$ across 64 scans and at the spectral resolution of 2 $cm^{-1}$.

Atomic Force Microscopy (AFM) was carried out on the 2D-0D hybrid with an AFM instrument MFP-3D (Asylum Research, Santa Barbara, CA, USA), using NCHR (NanoWorld, Neuchâtel, Switzerland) probes in tapping mode in air. The images collected were processed with the AFM company software IgorPro 6.22 (Wavemetrics, Lake Oswego, OR, USA).

*Spectroscopic characterization*: The ultrafast differential transmission measurements have been performed by using a Light Conversion Pharos and Coherent Libra amplified laser systems with the fundamental wavelength at 1030 and 800 nm, respectively, a pulse duration of about 200 fs and 100 fs, and a repetition rate of 100 kHz and 1 kHz, respectively. Noncollinear optical parametric amplifiers (NOPA) has been built with a procedure reported in Reference [16]. With the 100kHz setup the NOPA was tuned to have 60 fs pulses with a central wavelength of 1750 nm with a fluence of 300 μJ/$cm^2$, while with the 1 kHz setup we set the NOPA at 500nm.

For the 100kHz setup, white light generation for the probe pulse has been achieved focusing the fundamental beam into a 6mm YAG crystal. The differential transmission $\frac{\Delta T}{T} = \frac{(T_{ON}-T_{OFF})}{T_{OFF}}$ has been acquired using a common path interferometric spectrometer [17] followed by a Si photodiode, the signal was then acquired with a lock-in amplifier. The high repetition rate of the laser combined with a lock-in detection allowed to achieve high sensitivity.

For the other configuration with the laser at 1kHz the fundamental beam has been focused onto a sapphire plate of 2mm to generate the white light for the probe beam. The signal has been than acquired with an optical multichannel analyzer. In both configuration the probe was generated to cover both A and B exciton of the MoS2.

The Fermi level of ITO is below the bottom of the conduction band of $MoS_2$ [18]. In a heterojunction between ITO nanocrystals and $MoS_2$ monolayers (Figure 1a) it is possible to observe plasmon induced hot electron extraction via intraband excitation of ITO [5,19] (Figure 1b).

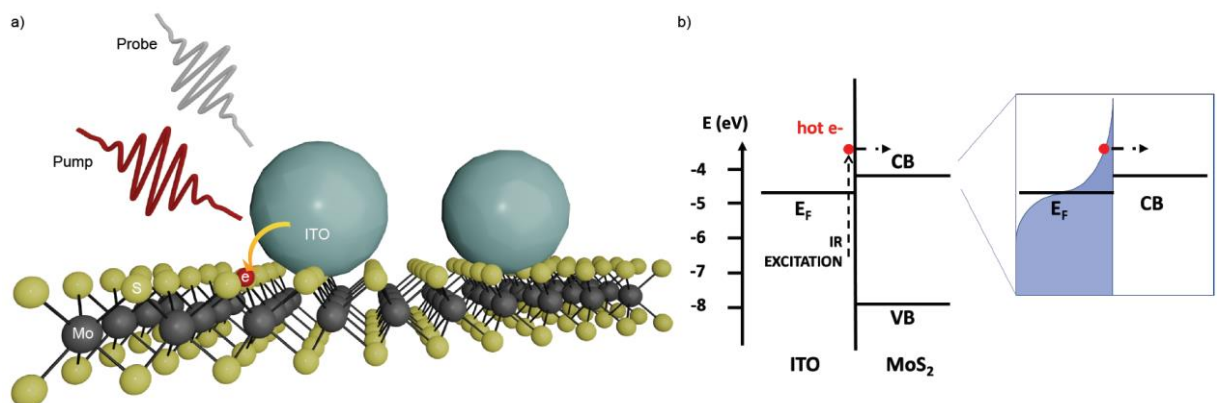

*Figure 1 a) ITO/MoS2 heterojunction sketch; b) Band alignment between ITO and MoS2, on the right hand side we have the sketch of the hot Fermi Dirac Distribution.*

Our sample consists of monolayer molybdenum disulphide (1L-$MoS_2$) covered with indium tin oxide nanocrystals (ITO NCs). The ITO nanocrystals have been synthesized following a synthesis protocol from [20]. Details are given in the methods section. The ITO NCs are approximately 13 nm in size (see Figure S1 in the Supporting Information) with a doping level of around 11 % (Sn/tot). The monolayers have been produced by gold assisted exfoliation, which resulted into the fabrication of large areas of monolayers in the range of hundreds of micrometers. The monolayer nature of the 1L-$MoS_2$ was identified by Raman and photoluminescence (PL) spectroscopy (Figure 2a and b). The hybrid structure was formed by spin coating the ITO NCs over 1L-$MoS_2$ on a Silica substrate, followed by annealing at 300 °C to remove the surfactants that are typically covering the ITO NCs and improve the film conductivity. An FTIR spectrum is given in Figure 2c before and after spin coating. The presence of the ITO NCs is identified by the near infrared peak in the hybrid sample. It corresponds to the localized surface plasmon resonance, which is due to the high doping level in the NCs. The broadening and red shift of the LSPR is a result of the deposition of the ITO NCs film on a substrate. A typical micrograph of the 1L-$MoS_2$ /ITO NCs hybrid is shown in Figure 2d, where the film of ITO NCs covers both the 1L-$MoS_2$ (purple areas) and the substrate (grey areas). Figure 2e shows an AFM micrograph of a similar area displaying the overall thickness of the sample in the range of 15 nm as a result of the ITO NCs covering the 1L-$MoS_2$.

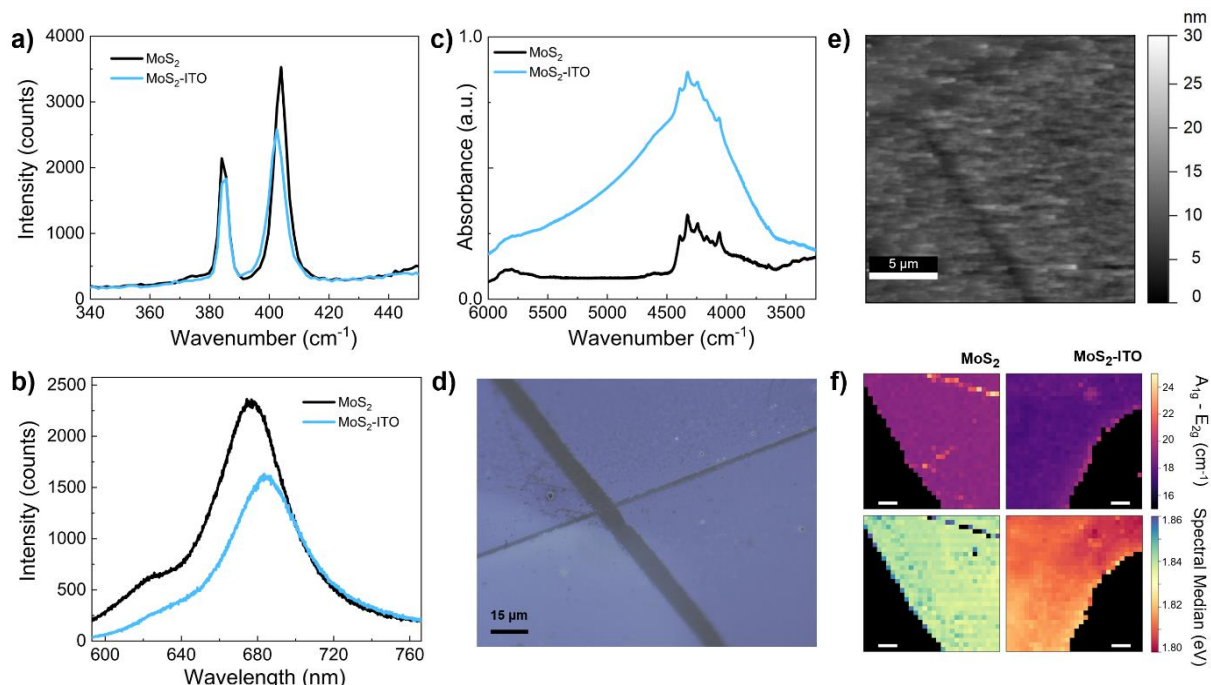

*Figure 2 a), b) and c) typical Raman, PL and NIR absorption spectra of 1L-MoS$_2$ (black curves) and hybrid 1L-MoS$_2$/ITO NCs (blue curves). d) Optical micrograph image displaying the 1L-MoS$_2$ covered in ITO NCs (scale bar is 15 µm) and e) AFM micrograph of a similar area (scale bar is 5 µm) highlighting the granular nature of the NC film with a thickness of ~15 nm. f) On top, Raman maps of the 1L-MoS$_2$ (left) and hybrid 1L-MoS$_2$/ITO NCs (right) reporting the separation between the $A_{1g}$ and $E_{2g}$ peaks. Below, PL maps of the same two areas showing the spectral median of the photoluminescence emission before (left) and after (right) the annealing process. Scale bar is 10 µm.*

We performed a Raman and photoluminescence map of the 1L-MoS$_2$ before and after depositing the ITO NCs to further investigate the spatial homogeneity of the two samples (Figure 2f). From the Raman maps no significant spatial inhomogeneities were detected. The monolayer nature of MoS$_2$ is obvious due to the separation of the $A_{1g}$ and $E^1_{2g}$ Raman peaks below 21 cm$^{-1}$ [21]. The post treatment of the hybrid sample upon annealing results into a small shift of the $A_{1g}$ peak of the 1L-MoS$_2$. The PL maps report the spectral median of the photoluminescence emission, showing again no significant spatial inhomogeneities in the two samples. Due to the heat treatment, the PL peak of the hybrid is systematically red-shifted with compared to the PL emission of the pristine 1L-MoS$_2$ sample. This is assigned to the n-type doping of the NCs upon annealing. A typical example PL spectrum of the 1L-MoS$_2$ and the hybrid is given in Figure 2b, highlighting the red shift and broadening of the spectrum as a result of additional trion PL from the 2D material doping. However, overall the results show that the sample remains intact after deposition and post treatment of the ITO NCs.

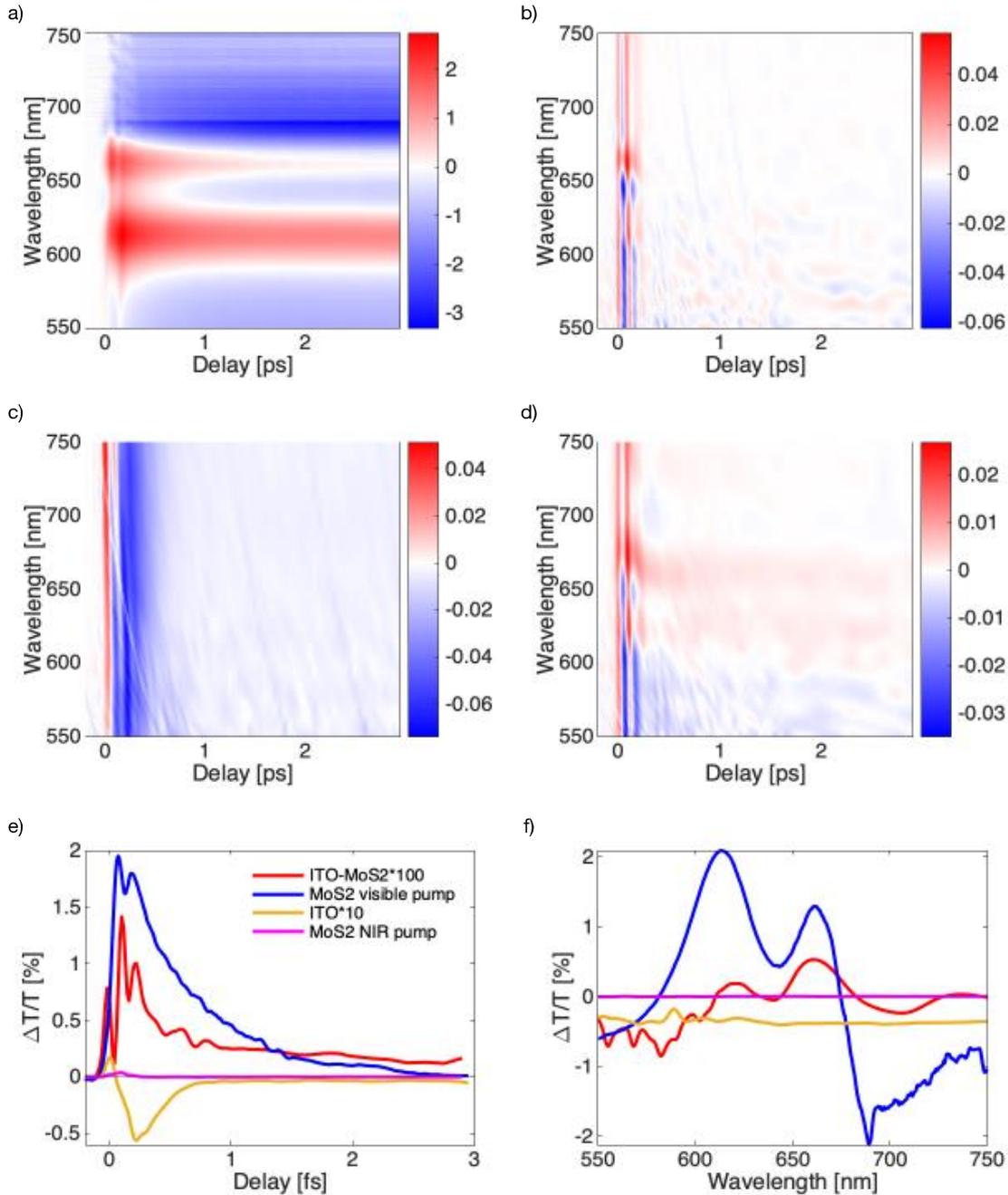

*Figure 3 a) MoS$_2$ excited above gap at 500nm b) MoS$_2$ excited at 1750 nm., c) ITO excited at 1750 nm, d) ITO-MoS$_2$ heterojunction excited at 1750 nm e) Kinetic of 663nm probe for all the sample, ITO-MoS$_2$ and ITO signal have been multiplied by 100 and 10 respectively to have a comparable signal with MoS$_2$. f) Spectra cut of the 4 different maps at a pump probe delay time of 350 fs.*

We perform broadband transient transmittivity to measure the non-equilibrium dynamics of the heterostructure. In figure 3 we reported the differential transmittivity map of the heterostructure with different excitation energy. In particular, in figure 3a) we show the result of monolayer MoS$_2$ pumped above optical bandgap at 500 nm (2.48 eV). The positive signals at ~630 nm and ~660 nm are ascribed to the bleaching of the excitonic transition induced by the pump due to the so called Pauly-blocking mechanism. Another effect is the renormalization of the excitonic binding energy and the electronic bandgap[22–24]. In panel b) we excited the MoS2 below bandgap at 1750 nm (0.7 eV), which is much lower with respect to the optical bandgap. We have chosen 1750 nm to ensure that even by two-photon absorption we could not reach the excitonic energy that is around 660 nm (1.87 eV). In this case, we can only see an artifact at zero-time delay between pump and probe, the so-called Cross phase modulation artifact (XPM). This is a well-known artifact that arise from the Kerr effect, when the intense pump pulse passes through a glass substrate modify the refractive index and induce a redistribution of the spectral component of the probe,[25,26] another artifact is related to the exciton, the optical stark effect that can be seen around 630 nm and 660 nm.[27] After those coherent artifacts, we do not have any remaining signal, confirming that we do not have any exchange of energy between $MoS_2$ and the pump pulse.

In figure 3c we can see the result from the excitation of an ITO nanocrystal film, After the excitation of the plasmon using the NOPA at 1750 nm, we can see a featureless negative signal in the whole visible spectrum. Following the excitation of the plasmon, in a timescale of about 10 fs [28] we have the dephasing and the generation of an out-of-equilibrium Fermi-Dirac (FD) distribution.[28] Through electron-electron scattering an equilibrium hot FD[29] is reached in few tens of fs. This energetic distribution will exchange energy by electron-phonon scattering that heat up the lattice, in plasmonic materials with common shapes this usually takes place in a ps timescale. A Fourth and much slower process that is the phonon-phonon scattering to cool down the lattice have a time scale of >10 ps[30].

A hot FD distribution results in a modification of the dielectric constant of the material. This has been modelled by Blemker et al. [31]. By changing the $\varepsilon_\infty$ we are changing the refractive index of the material and this increases the reflectivity of the ITO film. So, we decrease the transmission of the probe resulting in an overall negative featureless signal at every wavelength in the visible.

Finally, in figure 3d) we excited the heterojunction of ITO and $MoS_2$ at 1750 nm, now is clear that after the coherent artifact we have the fingerprint of the exciton around 630 and 660 nm, we assess this to an ultrafast plasmon-induced hot electron extraction. The high energy tail of the hot FD distribution has enough energy to overcome the Schottky barrier and jump to the $MoS_2$. When now the probe arrives, we have created a different dielectric environment for the $MoS_2$ having those charges in the valence band.

In figure 3e) we compared the kinetics of the 4 experiments at the same probe wavelength (663 nm), we changed the intensity of the bare ITO and ITO-$MoS_2$ to be similar to the only $MoS_2$ excited in the visible to better compare them. ITO shows an ultrafast negative recombination dynamic, faster than 1ps that arise from the electron-phonon interaction,[32] the $MoS_2$ excited in the visible shows a dynamic that reach zero around 3 ps, this fast dynamics have been attributed to: exciton-exciton annihilation[33] and exciton radiative recombination,[34] the heterojunction shows a positive dynamic lasting longer than 3 ps suggesting for a long lived charge extraction from the ITO nanocrystal. In figure 3d) we show the spectra of the different sample at a fixed pump probe delay of 350 fs, for both $MoS_2$ and ITO-$MoS_2$ we can recall the feature of the two exciton and their derivative shape, in particular in the case of the heterostructure the peaks seem to be red-shifted, this could be ascribed to the generation of more Trion and so to support the hypothesis of charge injection.

In conclusion, we have used ultrafast spectroscopy to see charge transfer process from Indium Tin Oxide to single layer $MoS_2$ heterojunction excited resonant to the plasmon at 1750 nm, well below the band gap of molybdenum disulphide. We ascribe this phenomenon to plasmon-induced hot electron extraction. The tunability of ITO plasmonic resonance in the NIR and the long living charge separation make this heterostructure an ideal candidate for light-harvesting application for low energy photon in the whole infrared spectra. This could be very interesting for the fabrication of infrared solar cells operating in the infrared based plasmon induced hot electron extraction.

**Acknowledgement**
This project has received funding from the European Research Council (ERC) under the European Union's Horizon 2020 research and innovation programme (grant agreement No. [816313]).

**References**

(1) Leenheer, A. J.; Narang, P.; Lewis, N. S.; Atwater, H. A. Solar Energy Conversion via Hot Electron Internal Photoemission in Metallic Nanostructures: Efficiency Estimates. *Journal of Applied Physics* **2014**, *115* (13), 134301. https://doi.org/10.1063/1.4870040.

(2) Wu, K.; Chen, J.; McBride, J. R.; Lian, T. Efficient Hot-Electron Transfer by a Plasmon-Induced Interfacial Charge-Transfer Transition. *Science* **2015**, *349* (6248), 632–635. https://doi.org/10.1126/science.aac5443.

(3) Reineck, P.; Brick, D.; Mulvaney, P.; Bach, U. Plasmonic Hot Electron Solar Cells: The Effect of Nanoparticle Size on Quantum Efficiency. *J. Phys. Chem. Lett.* **2016**, *7* (20), 4137–4141. https://doi.org/10.1021/acs.jpclett.6b01884.

(4) Tagliabue, G.; DuChene, J. S.; Abdellah, M.; Habib, A.; Gosztola, D. J.; Hattori, Y.; Cheng, W.-H.; Zheng, K.; Canton, S. E.; Sundararaman, R.; Sá, J.; Atwater, H. A. Ultrafast Hot-Hole Injection Modifies Hot-Electron Dynamics in Au/p-GaN Heterostructures. *Nat. Mater.* **2020**, *19* (12), 1312–1318. https://doi.org/10.1038/s41563-020-0737-1.

(5) Clavero, C. Plasmon-Induced Hot-Electron Generation at Nanoparticle/Metal-Oxide Interfaces for Photovoltaic and Photocatalytic Devices. *Nature Photon* **2014**, *8* (2), 95–103. https://doi.org/10.1038/nphoton.2013.238.

(6) Marangi, F.; Lombardo, M.; Villa, A.; Scotognella, F. (INVITED) New Strategies for Solar Cells Beyond the Visible Spectral Range. *Optical Materials: X* **2021**, *11*, 100083. https://doi.org/10.1016/j.omx.2021.100083.

(7) Lounis, S. D.; Runnerstrom, E. L.; Llordés, A.; Milliron, D. J. Defect Chemistry and Plasmon Physics of Colloidal Metal Oxide Nanocrystals. *J. Phys. Chem. Lett.* **2014**, *5* (9), 1564–1574. https://doi.org/10.1021/jz500440e.

(8) Kriegel, I.; Scotognella, F.; Manna, L. Plasmonic Doped Semiconductor Nanocrystals: Properties, Fabrication, Applications and Perspectives. *Physics Reports* **2017**, *674*, 1–52. https://doi.org/10.1016/j.physrep.2017.01.003.

(9) Ghini, M.; Curreli, N.; Lodi, M. B.; Petrini, N.; Wang, M.; Prato, M.; Fanti, A.; Manna, L.; Kriegel, I. Control of Electronic Band Profiles through Depletion Layer Engineering in Core–Shell Nanocrystals. *Nat Commun* **2022**, *13* (1), 537. https://doi.org/10.1038/s41467-022-28140-y.

(10) Lounis, S. D.; Runnerstrom, E. L.; Bergerud, A.; Nordlund, D.; Milliron, D. J. Influence of Dopant Distribution on the Plasmonic Properties of Indium Tin Oxide Nanocrystals. *J. Am. Chem. Soc.* **2014**, *136* (19), 7110–7116. https://doi.org/10.1021/ja502541z.

(11) Crockett, B. M.; Jansons, A. W.; Koskela, K. M.; Johnson, D. W.; Hutchison, J. E. Radial Dopant Placement for Tuning Plasmonic Properties in Metal Oxide Nanocrystals. *ACS Nano* **2017**, *11* (8), 7719–7728. https://doi.org/10.1021/acsnano.7b01053.

(12) Tandon, B.; Ghosh, S.; Milliron, D. J. Dopant Selection Strategy for High-Quality Factor Localized Surface Plasmon Resonance from Doped Metal Oxide Nanocrystals. *Chem. Mater.* **2019**, *31* (18), 7752–7760. https://doi.org/10.1021/acs.chemmater.9b02917.

(13) Sakamoto, M.; Kawawaki, T.; Kimura, M.; Yoshinaga, T.; Vequizo, J. J. M.; Matsunaga, H.; Ranasinghe, C. S. K.; Yamakata, A.; Matsuzaki, H.; Furube, A.; Teranishi, T. Clear and Transparent Nanocrystals for Infrared-Responsive Carrier Transfer. *Nat Commun* **2019**, *10* (1), 1–7. https://doi.org/10.1038/s41467-018-08226-2.

(14) Zhou, D.; Li, X.; Zhou, Q.; Zhu, H. Infrared Driven Hot Electron Generation and Transfer from Non-Noble Metal Plasmonic Nanocrystals. *Nat Commun* **2020**, *11* (1), 2944. https://doi.org/10.1038/s41467-020-16833-1.

(15) Liu, F.; Wu, W.; Bai, Y.; Chae, S. H.; Li, Q.; Wang, J.; Hone, J.; Zhu, X.-Y. Disassembling 2D van Der Waals Crystals into Macroscopic Monolayers and Reassembling into Artificial Lattices. *Science* **2020**, *367* (6480), 903–906. https://doi.org/10.1126/science.aba1416.

(16) Cerullo, G.; Manzoni, C.; Lüer, L.; Polli, D. Time-Resolved Methods in Biophysics. 4. Broadband Pump–Probe Spectroscopy System with Sub-20 Fs Temporal Resolution for the Study of Energy Transfer Processes in Photosynthesis. *Photochem. Photobiol. Sci.* **2007**, *6* (2), 135–144. https://doi.org/10.1039/B606949E.

(17) Brida, D.; Manzoni, C.; Cerullo, G. Phase-Locked Pulses for Two-Dimensional Spectroscopy by a Birefringent Delay Line. *Opt. Lett., OL* **2012**, *37* (15), 3027–3029. https://doi.org/10.1364/OL.37.003027.

(18) Kriegel, I.; Ghini, M.; Bellani, S.; Zhang, K.; Jansons, A. W.; Crockett, B. M.; Koskela, K. M.; Barnard, E. S.; Penzo, E.; Hutchison, J. E.; Robinson, J. A.; Manna, L.; Borys, N. J.; Schuck, P. J. Light-Driven Permanent Charge Separation across a Hybrid Zero-Dimensional/Two-Dimensional Interface. *J. Phys. Chem. C* **2020**, *124* (14), 8000–8007. https://doi.org/10.1021/acs.jpcc.0c01147.

(19) Christopher, P.; Moskovits, M. Hot Charge Carrier Transmission from Plasmonic Nanostructures. *Annual Review of Physical Chemistry* **2017**, *68* (1), 379–398. https://doi.org/10.1146/annurev-physchem-052516-044948.

(20) Jansons, A. W.; Hutchison, J. E. Continuous Growth of Metal Oxide Nanocrystals: Enhanced Control of Nanocrystal Size and Radial Dopant Distribution. *ACS Nano* **2016**, *10* (7), 6942–6951. https://doi.org/10.1021/acsnano.6b02796.

(21) Niu, Y.; Gonzalez-Abad, S.; Frisenda, R.; Marauhn, P.; Drüppel, M.; Gant, P.; Schmidt, R.; Taghavi, N.; Barcons, D.; Molina-Mendoza, A.; de Vasconcellos, S.; Bratschitsch, R.; Perez De Lara, D.; Rohlfing, M.; Castellanos-Gomez, A. Thickness-Dependent Differential Reflectance Spectra of Monolayer and Few-Layer MoS2, MoSe2, WS2 and WSe2. *Nanomaterials* **2018**, *8* (9), 725. https://doi.org/10.3390/nano8090725.

(22) Pogna, E. A. A.; Marsili, M.; De Fazio, D.; Dal Conte, S.; Manzoni, C.; Sangalli, D.; Yoon, D.; Lombardo, A.; Ferrari, A. C.; Marini, A.; Cerullo, G.; Prezzi, D. Photo-Induced Bandgap Renormalization Governs the Ultrafast Response of Single-Layer MoS2. *ACS Nano* **2016**, *10* (1), 1182–1188. https://doi.org/10.1021/acsnano.5b06488.

(23) Borzda, T.; Gadermaier, C.; Vujicic, N.; Topolovsek, P.; Borovsak, M.; Mertelj, T.; Viola, D.; Manzoni, C.; Pogna, E. A. A.; Brida, D.; Antognazza, M. R.; Scotognella, F.; Lanzani, G.; Cerullo, G.; Mihailovic, D. Charge Photogeneration in Few-Layer MoS2. *Advanced Functional Materials* **2015**, *25* (22), 3351–3358. https://doi.org/10.1002/adfm.201500709.

(24) Trovatello, C.; Katsch, F.; Borys, N. J.; Selig, M.; Yao, K.; Borrego-Varillas, R.; Scotognella, F.; Kriegel, I.; Yan, A.; Zettl, A.; Schuck, P. J.; Knorr, A.; Cerullo, G.; Conte, S. D. The Ultrafast Onset of Exciton Formation in 2D Semiconductors. *Nature Communications* **2020**, *11* (1), 5277. https://doi.org/10.1038/s41467-020-18835-5.

(25) Lorenc, M.; Ziolek, M.; Naskrecki, R.; Karolczak, J.; Kubicki, J.; Maciejewski, A. Artifacts in Femtosecond Transient Absorption Spectroscopy. *Applied Physics B: Lasers and Optics* **2002**, *74* (1), 19–27. https://doi.org/10.1007/s003400100750.

(26) Ekvall, K.; van der Meulen, P.; Dhollande, C.; Berg, L.-E.; Pommeret, S.; Naskrecki, R.; Mialocq, J.-C. Cross Phase Modulation Artifact in Liquid Phase Transient Absorption Spectroscopy. *Journal of Applied Physics* **2000**, *87* (5), 2340–2352. https://doi.org/10.1063/1.372185.

(27) Cunningham, P. D.; Hanbicki, A. T.; Reinecke, T. L.; McCreary, K. M.; Jonker, B. T. Resonant Optical Stark Effect in Monolayer WS2. *Nat Commun* **2019**, *10* (1), 5539. https://doi.org/10.1038/s41467-019-13501-x.

(28) Besteiro, L. V.; Yu, P.; Wang, Z.; Holleitner, A. W.; Hartland, G. V.; Wiederrecht, G. P.; Govorov, A. O. The Fast and the Furious: Ultrafast Hot Electrons in Plasmonic Metastructures. Size and Structure Matter. *Nano Today* **2019**, *27*, 120–145. https://doi.org/10.1016/j.nantod.2019.05.006.

(29) Hartland, G. V.; Besteiro, L. V.; Johns, P.; Govorov, A. O. What's so Hot about Electrons in Metal Nanoparticles? *ACS Energy Lett.* **2017**, *2* (7), 1641–1653. https://doi.org/10.1021/acsenergylett.7b00333.

(30) Hu, M.; Chen, J.; Li, Z.-Y.; Au, L.; Hartland, G. V.; Li, X.; Marquez, M.; Xia, Y. Gold Nanostructures: Engineering Their Plasmonic Properties for Biomedical Applications. *Chem. Soc. Rev.* **2006**, *35* (11), 1084. https://doi.org/10.1039/b517615h.

(31) Blemker, M. A.; Gibbs, S. L.; Raulerson, E. K.; Milliron, D. J.; Roberts, S. T. Modulation of the Visible Absorption and Reflection Profiles of ITO Nanocrystal Thin Films by Plasmon Excitation. *ACS Photonics* **2020**, *7* (5), 1188–1196. https://doi.org/10.1021/acsphotonics.9b01825.

(32) Guo, P.; Schaller, R. D.; Ketterson, J. B.; Chang, R. P. H. Ultrafast Switching of Tunable Infrared Plasmons in Indium Tin Oxide Nanorod Arrays with Large Absolute Amplitude. *Nature Photon* **2016**, *10* (4), 267–273. https://doi.org/10.1038/nphoton.2016.14.

(33) Sun, D.; Rao, Y.; Reider, G. A.; Chen, G.; You, Y.; Brézin, L.; Harutyunyan, A. R.; Heinz, T. F. Observation of Rapid Exciton–Exciton Annihilation in Monolayer Molybdenum Disulfide. *Nano Lett.* **2014**, *14* (10), 5625–5629. https://doi.org/10.1021/nl5021975.

(34) Robert, C.; Lagarde, D.; Cadiz, F.; Wang, G.; Lassagne, B.; Amand, T.; Balocchi, A.; Renucci, P.; Tongay, S.; Urbaszek, B.; Marie, X. Exciton Radiative Lifetime in Transition Metal Dichalcogenide Monolayers. *Phys. Rev. B* **2016**, *93* (20), 1–10. https://doi.org/10.1103/PhysRevB.93.205423.

**Supporting Information**

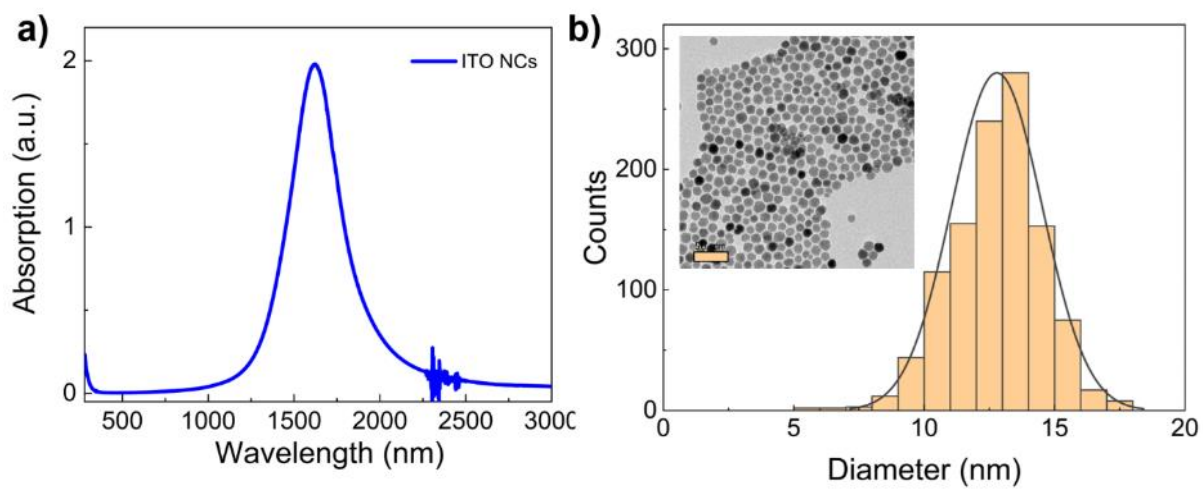

*Figure S1 a) Absorption spectrum of the ITO NCs in colloidal solution. The localized surface plasmon resonance dominates the absorption in the NIR region. b) size distribution of the ITO NCs with typical TEM image (inset). Scale bar is 50 nm.*